\documentstyle[12pt]{article}
\begin{document}
\baselineskip 6mm
\thispagestyle{empty}
\begin{flushright}
  {\tt DPNU-95-36 \\ October 1995}
\end{flushright}
 \vspace{0.1em}
\vspace*{5mm}
\begin{center}
   {\Large\bf New $I=J$ Rules for the Baryon}
\vspace{3mm} \\
   {\Large\bf Vertices in $1/N_c$ Expansion}
\vspace{13mm} \\
 {\large\sc Akira Takamura}
  \footnote[1]{
                e-mail addresses:
                takamura@eken.phys.nagoya-u.ac.jp
              }
\vspace{3mm} \\
  \sl{Department of Physics, Nagoya University,}\\
  \it{Nagoya 464-01, Japan}

\vspace{30mm}  

\center{\bf Abstract} \\
\vspace{3mm}
\begin{minipage}[t]{120mm}
\baselineskip 5mm
{\small
 We apply the $1/N_c$ expansion in QCD to the baryon vertices.
 We find new model independent properties for the isoscalar
 and isovector baryon vertices from the view point of $1/N_c$
 expansion in QCD.
 One of these results, $I=J$ rule, have been already found.
 The other properties, new $I=J$ rules, are the rules about
 the isospin and strangeness dependence
 for the baryon vertices.
}

\end{minipage}
\end{center}



\newpage
\baselineskip 18pt
\parskip 3pt
\section{Introduction}
\hspace*{\parindent}
 The $1/N_c$ expansion was proposed as a non-perturbative
 approach to QCD by 't~Hooft in 1974$\cite{tHooft}$.
 He has shown that the Feynman diagrams which are relevant
 at the leading order of the $1/N_c$ expansion are the
 planar diagrams without internal quark loops.
 Based on this $1/N_c$ expansion Witten suggested
 that in the large $N_c$ limit a baryon looks like
 a soliton$\cite{Witten}$.
 {}From this viewpoint the Skyrme's conjecture that baryons are
 solitons of the nonlinear chiral Lagrangian
 for the chiral fields has been revived in early 1980's and
 has succeeded in describing the baryon sector from the meson
 sector at least semi-quantitatively$\cite{Adkins}$.
 \par
 The $1/N_c$ expansion method has been considered to be
 a qualitative method to study QCD.
 However recent extensive studies of the consistency conditions
 $\cite{Gervais}\cite{Dashen}$ show that the $1/N_c$ expansion
 method is useful for obtaining also quantitative results of QCD
 from the model independent viewpoint.
 \par
 In the previous papers$\cite{Takamura}$ we have calculated
 the $F/D$ ratios of flavor $SU(3)$ symmetry for both
 spin-flip and spin-nonflip baryon vertices
 in the non-relativistic quark model(NRQM)
 and the chiral soliton model(CSM) for arbitrary color degrees
 of freedom $N_c$.
 The values of $F/D$ ratios for the spin-flip and spin-nonflip
 baryon vertices tends to $1/3$ and $-1$, respectively
 in the large $N_c$ limit.
 The physical meaning of these values turned out to be nothing
 but the $I=J$ rule indicated by Mattis and collaborators
 \cite{Mattis}, namely, the isovector dominance
 for the spin-flip baryon vertex and the isoscalar dominance
 for the spin-nonflip baryon vertex.
 \par
 In this paper we find new and model independent properties
 for the isoscalar and isovector baryon vertices
 from the viewpoint of $1/N_c$ expansion in QCD.
 There new properties are the rules about isospin
 and strangeness dependence for the baryon vertices.
 \par
 In section 2, we define the baryon states for arbitrary $N_c$
 and explain the $F/D$ ratios for the baryon vertices.
 In section 3, we will derive the isoscalar and isovector
 formulas for the baryon vertices with $N_c$.
 In section 4, we will find new $I=J$ rules by using the $F/D$
 ratios derived from the non-relativistic quark model
 and the chiral soliton model.


\section{The $N_c$ Dependence of Baryon States and $F/D$ Ratios}
\hspace*{\parindent}
 In this section we investigate the $N_c$ dependence
 for the baryon vertices from the flavor information
 in the $SU(N_c)$ QCD.
 {}From now on we assume the flavor $SU(3)$ symmetry
 in order to make arguments easy.
 \par
 In order to study the  properties of baryons in the $SU(N_c)$
 symmetric QCD with arbitrary $N_c$ we have to introduce
 the extended baryon state which is totally antisymmetric
 color singlet state of the $SU(N_c)$ symmetry.
 There are some ambiguities in extending the baryons
 for large $N_c$ and we have to introduce some unphysical
 members of the $SU(N_f)$ multiplet.
 Therefore we need to fix the ``physical states" of baryon
 in the $SU(N_c)$ QCD.
 We will show that the following special choice of
 extension for the large $N_c$ baryon is appropriate to obtain
 the correct large $N_c$ behavior of various baryon
 matrix elements.
 \par
 The ground state spin 1/2 baryons for arbitrary  $N_c$
 belong to $(k+1)(k+3)$ dimensional representation of
 flavor $SU(3)$ symmetry because of total symmetry in the
 spin-flavor states where $k = (N_c -1)/2$ $(k=0,1,2,\cdots)$.
 This representation is specified by the
 Young diagram with the first row of length $k+1$ and the
 second row of length $k$ and the root diagram.
 The physical octet baryons are located at the top region of
 this root diagram.
 The flavor wave functions are represented by
 the tensors with one superscript and $k$ subscripts and
 the physical octet baryons are given by $\cite{Dashen}$.
 The hypercharge extended to arbitrary $N_c$ is given by
\begin{equation}
Y= \frac{N_c B}{3} + S,
\end{equation}
 where $B$ is the baryon number and $S$ is the
 strangeness which reduces to $Y=B+S$
 in the physical case of $N_c=3$.
 \par
 The most general flavor octet vertex is represented
 as a sum of two independent terms as follows:
\begin{equation}
    <B_f \mid O^{a} \mid B_i>
= F\;{\rm tr}(\lambda^a [{\bar B}_f,B_i]) + D\;{\rm tr}
    (\lambda^a \{{\bar B}_f,B_i\}), \label{FD}
\end{equation}
 where $\lambda^a$ is  a flavor octet matrix with $a=1,..., 8$.
 \par
 For the spin-nonflip vertex $<B_f|O^a|B_i>$,
 this $F/D$ ratio will be denoted by $F_+/D_+$.
 For the spin-flip vertex $<B_f|O^{ai}|B_i>$ we have
 a similar expression and the $F/D$ ratio is denoted
 by $F_-/D_-$.
 \par
 The $F/D$ ratios express the spin structure of some specific
 model, for instance NRQM and CSM are useful to observe
 the large $N_c$-dependence of QCD, independently of details of
 the specific effective model of QCD.
 \par
 In the NRQM and CSM we calculate the $F/D$ ratios
 \cite{Takamura}
\begin{eqnarray}
& & \left(\frac{F_+}{D_+}\right)_{SU(6)~NRQM}
 =  -\frac{N_c+1}{N_c-3}
 =  -1 - \frac{4}{N_c} + \frac{12}{N_c^2} + \cdots,
  \label{fdnrqm} \\
& & \left(\frac{F_+}{D_+}\right)_{SU(3)~CSM}
 =  -\frac{N_c^2+4N_c-1}{N_c^2+4N_c-9}
 =  -1 + \frac{0}{N_c} - \frac{8}{N_c^2} + \cdots,
  \label{fdcsm}
\end{eqnarray}
Similarly,
\begin{eqnarray}
& & \left(\frac{F_-}{D_-}\right)_{SU(6)~NRQM}
 =  \frac{N_c+5}{3(N_c+1)}
 =  \frac{1}{3} + \frac{4}{3N_c} - \frac{4}{3N_c^2} + \cdots,
  \label{fdnrqm1} \\
& & \left(\frac{F_-}{D_-}\right)_{SU(3)~CSM}
 =  \frac{N_c^2 + 8N_c +27}{3(N_c^2 + 8N_c +3)}
 =  \frac{1}{3} + \frac{0}{N_c} + \frac{8}{N_c^2} + \cdots,
    \label{fdcsm1}
\end{eqnarray}
 These calculation indicate that structure of baryon vertices
 become the same for both the NRQM and CSM exist
 in the large $N_c$ limit\cite{Manohar}.
 Furthermore in the CSM the $1/N_c$ correction dose not exist.


\section{The Model Independent Analysis for the
 Baryon Vertices in $1/N_c$ Expansion}
\hspace*{\parindent}
 We calculate isoscalar and isovector baryon vertices
 in order to derive the $I=J$ rules from the baryon wave
 function which we have considered in the previous section.
 \par
 According to the Wigner-Eckart theorem, the matrix element of
 the diagonal operator $H_8$ can generally be expressed as
 follows
\begin{eqnarray}
& & <B_f \mid H_8 \mid B_i> \nonumber \\
& & =- F + \frac{1}{3}D + \frac{2}{k}(F-D) K \nonumber \\
& & + \frac{1}{k+2}
     \left(
            F+D - \frac{1}{k}(F-D)
     \right)
     \left(
           I(I+1)-(K+\frac{1}{2})
           (K+\frac{3}{2})
     \right), \label{isoscalar}
\end{eqnarray}
where $K$ is related to hypercharge or strangeness by
 $Y=N_c/3B+S=N_c/3B-2K$.
 \par
 We find two different structures from this formula.
 One is operator structure and the other is a special
 pattern of coefficients.
 \par
 The operator structure consists of three parts.
 The first line of eq.(\ref{isoscalar}) gives the same
 contribution to all states $-F+1/3D$ and the increasing
 with $K$ contribution.
 The second line of eq.(\ref{isoscalar})
 appears only for the states located on the inner
 triangle of the root diagram because of the factor
  $I(I+1) - (K+1/2)(K+3/2)$.
 \par
 Another structure of this formula is that
 the combination of $F$ and $D$ appear in only three
 patterns, $-F+1/3D, F-D, F+D$.
 The coefficients $-F+1/3D, F+D$ are related to the
 limiting value $F_+/D_+=-1$ and $F_-/D_-=1/3$, respectively.
 \par
 This formula does not depend on the extrapolation while
 the $N_c$ counting of the operators $K$ and
 $I(I+1) - (K+1/2)(K+3/2)$ depend on the extrapolation.
 {}From the previous argument it is natural that isospin and
 strangeness are $O(N_c^0)$.
 \par
 If we identify the ``physical states" to the states which have
 the same spin and isospin, the Okubo-Gell-Mann mass relation
\begin{equation}
3\Lambda + \Sigma = 2(N + \Xi) , \label{og}
\end{equation}
 holds for the arbitrary number $N_c$.
 \par
 Next we consider isovector diagonal matrix elements.
 After the tedious calculation the diagonal matrix element
 of operator $H_3$ can be expressed by using the Wigner 6j
 symbol as follows.
\begin{eqnarray}
& & < I,I_3,K \mid H_3 \mid I,I_3,K > \nonumber \\
& & =
    (-1)^{3/2-I-K} \sqrt{2{\rm dim}I}
   \left\{
      \begin{array}{ccc}
      1 & I           & I           \\
      K & \frac{1}{2} & \frac{1}{2} \\
      \end{array}
   \right\}
   \left(
      \begin{array}{cc|c}
      I   & 1 & I   \\
      I_3 & 0 & I_3 \\
      \end{array}
   \right)
      \left[
             F+D + \frac{2}{k}(F-D)K
      \right.
          \nonumber \\
& & + \frac{3}{k+2}
    \left.
      \left(
          -F + \frac{1}{3}D - \frac{1}{k}(F-D)
      \right)
      \left(
           I(I+1)-(K+\frac{1}{2})(K+\frac{3}{2})
      \right)
    \right] \label{isovector}
\end{eqnarray}
 This isovector formula has the same operator structure
 are the isoscalar formula (\ref{isoscalar}).
 The common contribution is replaced by $F+D$ instead of
 $-F+1/3D$ in the isospin nonflip vertex.
 The Clebsh-Goldan coefficient express as the isospin
 conservation.
 \par
 The off-diagonal matrix element of $H_3$ is given by
\begin{eqnarray}
& & < I-1,I_3,K \mid H_3 \mid I,I_3,K > \nonumber \\
& & = < I,I_3,K \mid H_3 \mid I-1,I_3,K > \nonumber \\
& & =
    (-1)^{5/2-I-K} \sqrt{2{\rm dim}I}
   \left\{
      \begin{array}{ccc}
      1 & I           & I-1         \\
      K & \frac{1}{2} & \frac{1}{2} \\
      \end{array}
   \right\}
   \left(
      \begin{array}{cc|c}
      I   & 1 & I-1 \\
      I_3 & 0 & I_3 \\
      \end{array}
   \right) \nonumber \\
& &  \times \sqrt{\frac{k-2K+1}{k+2}}
   \left(
          F+D-\frac{1}{k}(F-D)
   \right) \label{isovector1}
\end{eqnarray}
 This result is consistent with the result given by Dashen,
 Jenkins and Manohar\cite{Dashen}
\begin{eqnarray}
& & < I',I'_3,J',J'_3;K \mid A^{ia} \mid I,I_3,J,J_3;K >
    \nonumber \\
& & =
    N_c g(K)(-1)^{2J'+j-I'-K} \sqrt{{\rm dim}I{\rm dim}J}
    \nonumber \\
& & \left\{
      \begin{array}{ccc}
      1 & I & I' \\
      K & J & J' \\
      \end{array}
   \right\}
   \left(
      \begin{array}{cc|c}
      I   & 1 & I'   \\
      I_3 & a & I'_3 \\
      \end{array}
   \right)
   \left(
      \begin{array}{cc|c}
      J   & 1 & J'   \\
      J_3 & i & J'_3 \\
      \end{array}
   \right)
\end{eqnarray}
 The Coleman-Glashow mass relation
\begin{eqnarray}
\Sigma^+ - \Sigma^- = p -n + \Xi^0 - \Xi^-
\end{eqnarray}
 is obtained by assuming the same extrapolation with isoscalar
 part only for $N_c = 3$ contrary to the case of Okubo-Gell-Mann
 mass relation.


\section{New $I=J$ Rules for the Baryon Vertices in $1/N_c$
 Expansion}
\par
\hspace*{\parindent}
 In the previous section we have constructed
 the general isoscalar and isovector baryon vertices
 in terms of the Wigner-Eckart theorem.
 However we were not able to decide the $N_c$ dependence only
 by use of information on flavor.
 In order to decide the $N_c$ dependence completely,
 we need to know the values of $F/D$ ratios.
 The $F/D$ ratios is related to the spin dependence of
 the system.
 \par
 If the spin-nonflip part of $H_8$ transforms as
 $(SU(3)_f,SU(2)_s)={\bf (8,1)}$, then
 $F_+/D_+ = -1 + O(1/N_c)$.
 With this $F_+/D_+$ ratio for the mass formula for the
 baryons, the mass difference of baryons is given by
\begin{equation}
    H(I=J=0)
 =  N_c a + b K + \frac{c}{N_c}
    \left(
          { I(I+1)-(K+\frac{1}{2})(K+\frac{3}{2})}
    \right), \label{I=J=0}
\end{equation}
where $a$, $b$ and $c$ are $O(N_c^0)$.
 \par
 The isovector formula eq.(\ref{isovector}) is
\begin{equation}
   H^a(I=1,J=0)
 = \left[
         d' + e' K + f'
   \left(
          I(I+1)-(K+\frac{1}{2})(K+\frac{3}{2})
   \right)
   \right] I^a, \label{I=1,J=0}
\end{equation}
 We find that the terms which depend on the isospin and
 the strangeness are suppressed to the order $O(1/N_c)$.
 A similar formula is also derived in Ref.{\cite{Dashen}}
 \par
 On the other hand if the spin-flip part of $H_8$ transforms as
  $(SU(3)_f,SU(2)_s)={\bf (8,3)}$, then $F_-/D_- = 1/3
  + O(1/N_c)$. We can apply this formula
 to the isoscalar part of the magnetic moment.
 Then we obtain
\begin{equation}
   H^i(I=0,J=1)
 = \left[
          a' + b'  K
         + c'
   \left(
          I(I+1)-(K+\frac{1}{2})(K+\frac{3}{2})
   \right)
   \right] J^i \label{I=0,J=1}
\end{equation}
where $a'$, $b'$ and $c'$ are $O(N_c^0)$.
 \par
 The isovector formula which we can apply
 to the isovector part of the magnetic moment is
\begin{equation}
   H^{ia}(I=J=1)
 = \left[
         N_c d + e K + \frac{f}{N_c}
   \left(
          I(I+1)-(K+\frac{1}{2})(K+\frac{3}{2})
   \right)
   \right] X_0^{ia} \label{I=J=1}
\end{equation}
 {}From these results we find that there are contributions
 with two different properties.
 \par
 One is the old $I=J$ rule found by Mattis and
 collaborators \cite{Mattis} and the other is
 the new $I=J$ rule.
 The old $I=J$ rule is concerned with the leading order
 in $1/N_c$ expansion and new $I=J$ rule
 is related to the $N_c$ dependent part of isospin and
 strangeness.
 \par
 If spin and isospin of the octet operator are equal,
 then its baryon vertices are $O(N_c)$ from (\ref{I=J=0}) and
 (\ref{I=J=1}).
 But if spin and isospin are different, then these are
 terms of order $O(N_c^0)$ from (\ref{I=1,J=0}) and
 (\ref{I=0,J=1}).
 This property is the old $I=J$ rule.
 \par
 On the other hand we recognize new properties from these
 results. It's properties are the isospin and strangeness
 dependence of baryon vertices.
 The isospin and strangeness are suppressed by the $1/N_c$
 expansion if we use (\ref{I=J=0}) or (\ref{I=J=1}),but
 in (\ref{I=1,J=0}) and (\ref{I=0,J=1}) the isospin and
 strangeness dependence survives even if we take the limit
 $N_c \to \infty$. These properties belong to the
 new $I=J$ rules.
 \par
 The isospin and strangeness dependence are not necessarily
 to be suppressed at the lowest order in $1/N_c$
 expansion $\cite{Dashen}\cite{Luty}\cite{Carone}$.


\section{Summary}
\hspace*{\parindent}
 We have found the new $I=J$ rules for the baryon vertices in
 $1/N_c$ expansion. The assumption which we have used to derive
 the new $I=J$ rules are three, first is quark confinement,
 second is $SU(3)$ flavor symmetry and the last is that
 spin and isospin of baryon states in the $SU(N_c)$ QCD
 are $O(N_c^0)$.
 We guess that the second assumption is not essential from
 the consideration of consistency condition approach
 \cite{Dashen}.
 \par
 In our discussion there exists some unsatisfactory point.
 One problem is that we have used the special models in
 order to calculate the $F/D$ ratios.
 By virtue of this argument we cannot understand $I=J$ rules
 as easy as the OZI rule.
 The second problem is that we assume $SU(3)$ flavor symmetry.
 We do not need this assumption to derive the $I=J$ rules.

\section*{Acknowledgement}
 The author would like to thank S. Sawada, S. Kitakado
 and Y. Matsui for many useful advise and reading this
 manuscript.


\end{document}